\documentstyle[12pt]{article}
\renewcommand{\thesection}{\Roman{section}}

\begin{document}

\begin{flushright}
KOBE--FHD--93--03\\
March  1994\\
\end{flushright}
\vspace{2em}

\begin{center}
\renewcommand{\thefootnote}{\fnsymbol{footnote}}
{\large \bf Gravitational Shock Waves for Schwarzschild\\
            and Kerr Black Holes}\\
\vspace{4em}
Koichi HAYASHI${}^{1}$ and Toshiharu SAMURA${}^{2}$
          \footnote[2]{e-mail address: samura@cphys.cla.kobe-u.ac.jp}\\
\vspace{3em}
${}^{1}${\it Department of Mathematics and Physics,}\\
        {\it Faculty of Science and Technology,}\\
        {\it Kinki University, Higashi-osaka,}\\
		{\it  Osaka 577, Japan}\\
\vspace{2em}
${}^{2}${\it Graduate School of Science and Technology, }\\
        {\it Kobe University, Nada, Kobe 657, Japan}\\
\end{center}
\vspace{2em}

\begin{center}
\begin{abstract}
\baselineskip 12pt
The metrics of gravitational shock waves for a Schwarzschild black hole
in ordinary coordinates
and for a Kerr black hole in Boyer-Lindquist coordinates are derived.
The Kerr metric is discussed
for two cases: the case of a Kerr black hole moving parallel to
the rotational axis,
and moving perpendicular to the rotational axis.
Then, two properties from the derived metrics are investigated:
the shift of a null coordinate
and the refraction angle crossing the gravitational shock wave.
Astrophysical applications for these metrics are discussed in short.
\end {abstract}
\end{center}
\clearpage
\baselineskip 16pt

\setcounter{equation}{0}
\renewcommand{\theequation}{\arabic{section}.\arabic{equation}}
\section{Introduction}

The metrics of black holes which are derived from the vacuum solution
of Einstein's equation have
been investigated for many years \cite{Chandra83}.
But the space--time structure of {\it{moving}} black holes have not
been studied so much.
Recently, several papers are published for this problem.
In particular, the problem of gravitational waves which
are emitted when two black holes encounter, and of the scattering of
elementary particles
with Planck scale energies are discussed \cite{D'Eath92,'t Hooft87}.
In this paper, we investigate the metric when a black hole
(Schwarzschild or Kerr ) is moving at the limit of light velocity.

Let us consider the case of  a Schwarzschild black hole. When a black
hole is moving
relative to an observer far from the black hole, the metric derived for
the rest system
of  the black hole should be Lorentz transformed to that of observer's
system.
Generally, the metric of a black hole is not invariant for Lorentz
transformations,
so the transformed metric has a different structure from that for the black
hole
system.
If a black hole is moving at the limit of light velocity in the z--direction
(Lorentz $\gamma$ factor $\rightarrow \infty$), the metric is changed to
the following
form:
\begin{eqnarray}
\lim_{\gamma \to \infty} ds^2~\longrightarrow~-du~dv+dx^2+dy^2-A(\rho)
\end{eqnarray}
where we put $\rho^2=x^2+y^2$,
$u=t-z$, and $v=t+z$. It is clear from this form that this metric describes
a space--time
which has two flat half--spaces, $u>0$ and $u<0$, and the
axially symmetric plane gravitational wave concentrated on the interface $u=0$.
Moreover, the infinite Lorentz boost
has changed the algebraic type of the Weyl tensor from the type D
to the type N for the limiting metric (1.1). The corresponding curvature
tensor vanishes everywhere except at the surface $u=0$, where its nonzero
components
have singularities of $\delta (u)$. Such a metric is often called as "the
gravitational shock wave" \cite{D'Eath92,'t Hooft85,Novikov89}.

Aichelburg and Sexl \cite{Aichelburg71} have obtained the shock wave metric by
performing an infinite Lorentz transformation of the metric of a Schwarzschild
black hole for isotropic coordinates.
Their result is
\begin{eqnarray}
A \left(\rho \right)~=~4p~\log \rho^2.
\end{eqnarray}
Here, $p=\gamma M$ and only the
leading term in $\gamma$ is retained.
Also in this paper, we take into account only the leading order of $\gamma$ in
$A(\rho )$.

In Sec.II, we calculate the metric of a Schwarzschild black hole in ordinary
coordinates
when the infinite Lorentz transformation is performed.
Then, the limit of $\gamma \rightarrow \infty$
is carried out by the method of Loust\'{o} and S\'{a}nchez \cite{Lousto89}:
first, carry out a Lorentz boost with a finite $\gamma$. Next, integrate the
result
by $u$. Then, take the limit $\gamma \rightarrow \infty$, and finally,
differentiate it by $u$.
With these procedures,
the metric of the gravitational shock wave of the form of (1.1) is
obtained. In this method, it is not necessary to take
the limit of $M \rightarrow 0$ like  Aichelburg--Sexl,
so that we can examine the space--time structure near the region $\rho \sim
2M$,
the Schwarzschild horizon.

In Sec.III, we seek for a metric of a Kerr black hole in Boyer--Lindquist
coordinates
after taking the infinite Lorentz transformation in the same way.
We find that the metric
for the case of a black hole moving parallel to the rotational axis (z--axis)
is
different
from that for perpendicular motion.
In Sec.IIIa, the parallel case is investigated in detail. After obtaining the
metric
in general,
we ask for the metric for the special case of the maximally rotating black hole
$M \rightarrow a$, where $a$ is the angular momentum per unit mass.
In this case, the event horizon is at $\rho=a$, equal to the case of a static
black
hole.
On the other hand, the metric far from the horizon almost coincides with
the metric of Schwarzschild type given by Sec.II. In Sec.IIIb, the metric for
moving
in the
perpendicular direction (we take the $x$--direction) is calculated. To see
the maximal effect about a Kerr black hole, we restrict our discussion only on
the
equatorial plane (i.e. $z=0$). The point we wish to emphasize is that the
metric depends on the sign of $y$ (i.e. perpendicular to the moving
direction and the rotating direction). This dependence comes from
$g_{t\phi}$ component of the Kerr metric. The metric for $M \rightarrow a$
is also given as before.

Next, two properties of the obtained metrics are investigated.
There are general properties for the gravitational shock wave for the metric of
(1.1) \cite{'t Hooft85}, that is, the discontinuity and the refraction of a
geodesic.
For a geodesic which crosses the shock wave located at $u=0$,
the discontinuity $\Delta v$ at $u=0$ is given by (See Appendix A)
\begin{eqnarray}
\Delta v=- 2\left. A(\rho) \right|_{\rho=\rho_{0}}~~,
\end{eqnarray}
where $\rho_0$ is the value of $\rho$ when the geodesic reaches $u=0$.
And as illustrated in Fig.1, the refraction angle, $\Delta \phi$, is given by
\begin{eqnarray}
\tan \left( \Delta \phi \right)~=~ \frac{1}{2} \left. \frac{dA(\rho)}{d \rho}
 \right|_{\rho=\rho_{0}}~~.
\end{eqnarray}
These $\Delta v$ and $\Delta \phi$ are calculated for the metrics obtained in
Sec.II and Sec.III.
In particular, when a Kerr black hole moves in the $x$--direction, the
refraction
angle depends on
the sign of $y$. This is to be attributed to the
dragging effect of the Kerr black hole.

The gravitational shock wave metric is not only the academic interest of
theoretical
physics, but also is applicable to the actual astrophysics.
In fact, this metric was employed by D'Eath \cite{D'Eath92} to solve the
problem
of the scattering of two
ultra-relativistic black holes which are approaching in parallel.
He assumed that the radiation from a head--on collision of two black holes of
equal
masses
is fairly isotropic and concluded that the efficiently of the transformation of
the gravitational energy  into radiation is close to 25 \%.
On the other hand, the upper limit of the efficiency would be
found on the basis of Hawking's theorem and is equal to $\sim$ 25 \%
\cite{Penrose}.
It is very interesting that these two values coincide.

The gravitational shock waves have also attracted interests in recent years
since they
shed a light in modeling quantum scattering processes at very high energies,
where gravitational effects dominate \cite{'t Hooft87}.

In Sec.IV, we consider as an astrophysical application, the effect
which is analogous to that of the gravitational lens.
When there is a static supermassive black hole between an observer and
a source which is receding by a high
red shift, the refraction of light rays radiated from the source is calculated.
In that case, the metric of the gravitational shock wave is dealt with as the
first
order
approximation of the astrophysical body which is running at very large
velocity.
If the body moves with $\gamma \sim $10, the first order approximation is
adequate.
However, the most distant quasar observed at present has $\gamma \sim$ 3,
so the result is only approximate in nature.

Sec.V is devoted to conclusions and discussions.

We use the natural unit of $c=G=$1, throughout.


\setcounter{equation}{0}
\section{Gravitational Shock Waves for Schwarz-schild Black Holes}

We consider a Schwarzschild metric in ordinary coordinates \cite{Chandra83}:
\begin{eqnarray}
ds^2_{S} &=& -\left(1-2M/\bar{r} \right) d\bar{t}^2+
          \left(1-2M/\bar{r} \right)^{-1} d\bar{r}^2+
          \bar{r}^2 d\bar{\theta}^2 +
          \bar{r}^2 \sin ^2\bar{\theta} d\bar{\phi}^2 \nonumber \\
     &=& ds^2_M +\Delta S^2_S~~,
\end{eqnarray}
where
\begin{eqnarray}
ds_M^2 &=& -d\bar{t}^2 +d\bar{r}^2 +\bar{r}^2 d\bar{\theta}^2 +
                \bar{r}^2 \sin ^2\bar{\theta}~d\bar{\phi}^2~~, \nonumber\\
\Delta S^2_S &=& \left( 2M/\bar{r} \right) d\bar{t}^2 +
                    \left( 2M/\bar{r} \right) / \left(1-2M/\bar{r} \right)
					d \bar{r}^2 ~~. \nonumber
\end{eqnarray}
We set $\bar{r}^2=\bar{x}^2+\bar{y}^2+\bar{z}^2$ and $M$ is the mass
of the black hole.
$ds^2_{M}$ is the metric of the Minkowski space--time, and $\Delta S^2_{S}$
represents
the surplus term. Now, perform a Lorentz transformation to the system where
the black hole
is moving  in the z--direction with the velocity $v$ close to 1:
\begin{eqnarray}
t=\gamma \left( \bar{t}+v \bar{z} \right)~~,~z= \gamma \left(\bar{z}
+v \bar{t} \right)~~,~
x=\bar{x}~~,~y=\bar{y}~~,
\end{eqnarray}
where
\begin{eqnarray}
\gamma~=~(1-v^2)^{-1/2} ~~.\nonumber
\end{eqnarray}
As $ds^2_{M}$ of (2.1) is invariant for Lorentz transformations, only
$\Delta S^2_{S}$ is transformed according to (2.2).
Putting $u=t-z$, and $v=t+z$, old coordinates, $\bar{x}^{i}$s, are changed
as follows:
\begin{eqnarray}
&&\bar{t} \rightarrow \gamma u~~, \nonumber\\
&&\bar{z} \rightarrow -\gamma u~~, \nonumber\\
&&\bar{r}^2 \longrightarrow \gamma^2 u^2+\rho^2~~, \nonumber\\
&&d \bar{r} \longrightarrow \frac{\gamma u~du}{\sqrt{u^2+\tilde{\rho}^2}}
+O(1/\gamma)~~,
\end{eqnarray}
where
\begin{eqnarray}
\rho^2=x^2+y^2~~,~~\tilde{\rho}=\rho /\gamma~~. \nonumber
\end{eqnarray}
As mentioned in Sec.I $g_{uu}$ is calculated to order $\sim  O(\gamma)$.
$\Delta S^2_{S}$ is then given by
\begin{eqnarray}
\Delta S^2_S = 2p\left[ \frac{1}{\sqrt{u^2 +\tilde{\rho}^2}}
            +\frac{u^2}{\sqrt{\left(u^2+\tilde{\rho}^2 \right)^3} -2\tilde{M}
              \left( u^2+\tilde{\rho}^2\right) } \right] du^2~~,
\end{eqnarray}
where we set the energy $p=\gamma M$ and $\tilde{M}=M/\gamma$.

Taking the limit of $v \rightarrow 1$
(namely  $\gamma \rightarrow \infty$), $\Delta S^2_{S}$ behaves like
the metric
of the gravitational shock wave. To see it, consider the first term of r.h.s.
of (2.4).
The integration of this term by $u$ is given by:
\begin{eqnarray}
\int \frac{du}{\sqrt{u^2+\tilde{\rho}^2}}~=~
       \log \left( u+\sqrt{u^2+\tilde{\rho}^2} \right)~~,
\end{eqnarray}
Now the limit of $\gamma~\rightarrow~\infty$ is taken:
\begin{eqnarray}
\lim_{\gamma \to \infty}  \int \frac{du}{\sqrt{u^2+\tilde{\rho}^2}} ~
=~\theta \left( u \right) \log \left( 2u \right)+\theta \left( -u \right)
\left[ \log \tilde{ \rho}^2 - \log \left( -2u \right) \right]~~,
\end{eqnarray}
Differentiating by $u$, we obtain (with redundant terms removed):
\begin{eqnarray}
\frac{d}{du} \left[
      \lim_{\gamma \to \infty} \int \frac{du}{\sqrt{u^2+\tilde{\rho}^2}}
\right]~
 =~\frac{1}{|u|} - \log \rho^2~\delta \left( u \right)~~.
\end{eqnarray}
Similarly, the second term is calculated:
\begin{eqnarray}
&&\lefteqn{\frac{d}{du} \left[
      \lim_{\gamma \to \infty}
            \int \frac{ u^2~~~du}{\sqrt{\left(u^2+\tilde{\rho}^2 \right)^3}
                             -2 \tilde{M}\left( u^2+\tilde{\rho}^2\right)}
\right]}
							 ~\nonumber \\
 &&=\frac{1}{|u|} - \left[ \log \rho^2 -\frac{\rho}{2M} \pi
 +4 \sqrt{ \frac{\rho^2}{4M^2} -1}
    \tan^{-1}  \sqrt{\frac{\rho+2M}{\rho-2M}} \right]
  \delta \left( u \right)~~,\nonumber\\
\end{eqnarray}
where use has been made of the formula
\begin{eqnarray}
&&\int \frac{ u^2~~~du}{\sqrt{\left(u^2+\tilde{\rho}^2 \right)^3}
                             -2 \tilde{M}\left( u^2+\tilde{\rho}^2\right)
}~~~~~~~
							 \nonumber \\
&&~~~=~\frac{\rho}{2M} \tan ^{-1} \left( \frac{u}{\tilde{\rho}} \right)
 +\log \frac{u-\tilde{\rho}+\sqrt{u^2+\tilde{\rho}^2}}
            {u+\tilde{\rho}-\sqrt{u^2+\tilde{\rho}^2}} \nonumber\\
&&~~~ -2 \sqrt{ \frac{\rho^2}{4M^2} -1}
    \tan^{-1} \left[ \sqrt{\frac{\rho+2M}{\rho-2M}}
                  \frac{\sqrt{u^2+\tilde{\rho}^2}-\tilde{\rho}}{u} \right]~~.
\end{eqnarray}
(2-4) for $\gamma \rightarrow \infty$ is now given by
\begin{eqnarray}
\lefteqn{\lim_{\gamma \to \infty} \Delta S^2_S \longrightarrow} \nonumber\\
&&4p \left[\frac{1}{|u|} - \left[ \log \rho^2
        -\frac{\rho}{4M} \left(\pi-4 \sqrt{ 1-\frac{4M^2}{\rho^2} }
    \tan^{-1}  \sqrt{\frac{\rho+2M}{\rho-2M}} \right) \right]
  \delta \left( u \right) \right] du^2.\nonumber\\
\end{eqnarray}
Introducing a new coordinate
\begin{eqnarray}
dv'~=~dv-4p~du/|u|~~,
\end{eqnarray}
the gravitational shock wave metric for a Schwarzschild black hole
has the final form (dash on $v'$ is omitted),
\begin{eqnarray}
&&\lefteqn{\lim_{\gamma \to \infty} ds^2_S \longrightarrow
    -du~dv+dx^2+dy^2}\nonumber \\
&&-4p \left[ \log \rho^2  -\frac{\rho}{4M} \left(\pi-4 \sqrt{ 1-
\frac{4M^2}{\rho^2} }
    \tan^{-1}  \sqrt{\frac{\rho+2M}{\rho-2M}} \right) \right]
  \delta \left( u \right)  du^2~~.\nonumber\\
\end{eqnarray}
This is the exact result.

We now consider the nature of this metric in various limits. When $\rho \gg
2M$,
(2.12) is reduced to:
\begin{eqnarray}
&&\lim_{\gamma \to \infty} ds^2_S {\longrightarrow \atop \rho \gg 2M}
    -du~dv+dx^2+dy^2\nonumber \\
&&-4p\left[ \log \rho^2  +1-\frac{\pi}{2} \frac{M}{\rho}+\frac{4}{3}
\frac{M^2}{\rho^2}
       +O\left( \frac{M^3}{\rho^3} \right) \right]
  \delta \left( u \right)  du^2.
\end{eqnarray}
Further, we take the limit $M \rightarrow 0$ and execute a scale transformation
$\hat{x^{i}} =e~x^{i}$ (e: exponential). Then (2.13) is equal to
the Aichelburg--Sexl
metric (A--S metric hereafter).
As mentioned in Sec.I, the coefficient of $g_{uu}$ represents the shift
$\Delta v$ of $v$ crossing $u=0$. In Fig.2 the bold line shows the value of
$\Delta v$
near $\rho \sim M$. It is to be noted that $\Delta v =$0
at the event horizon $\rho = 2M$.

{}From (1.1), (1.4) and (2.12), we can calculate the refraction angle,
$\Delta \phi$,
of a null geodesic crossing the shock wave. The result is depicted in
Fig.3 by the bold line. It is to be noted that the refraction angle becomes
90${}^\circ$ for $\rho =2M$. This is natural because
the event horizon of a Schwarzschild black hole is at $\rho =2M$, and
the event horizon is invariant under the Lorentz transformation for
the z--direction.
The metric of (2.12) represents the space--time structure for all region of
 $\rho$ compared with A--S metric, which describes only for $\rho \gg M$.


\setcounter{equation}{0}
\section{Gravitational Shock Wave for Kerr Black Holes}

{}From above discussions, it is apparent that the gravitational shock wave
solution
can be calculated for
any sort of static black hole metrics. In this section we apply this method for
 a Kerr black hole.
We consider a kerr black hole in Boyer--Lindquist coordinates \cite{Chandra83}:
\begin{eqnarray}
ds^2_K &=& -\left(1-\frac{2M \bar{r}}{\Sigma} \right) d \bar{t}^2
      -\frac{4M \bar{r} a \sin^2 \bar{\theta}}{\Sigma} d \bar{t}~d \bar{\phi}
      +\frac{\Sigma}{\Delta} d \bar{r}^2 \nonumber \\
  &&+\Sigma~d \bar{\theta}^2
    +\left( \bar{r}^2+a^2+ \frac{2M \bar{r} a^2 \sin^2 \bar{\theta}}{\Sigma}
	\right)
      \sin^2 \bar{\theta}~d \bar{\phi}^2~~,
\end{eqnarray}
where
\begin{eqnarray}
&&\Delta~=~\bar{r}^2-2M \bar{r}+a^2~~, \nonumber \\
&&\Sigma~=~\bar{r}^2+a^2 \cos^2 \bar{\theta}~~. \nonumber
\end{eqnarray}
The metric (3.1) is divided into the Minkowski metric
$ds^2_M$ and the other parts:
\begin{eqnarray}
ds^2_K~=~ds_M^2+\Delta S^2_{tt}+\Delta S^2_{t \phi}+\Delta S^2_{rr}
        +\Delta S^2_{\theta \theta}+\Delta S^2_{\phi \phi}~~,
\end{eqnarray}
where
\begin{eqnarray}
\Delta S^2_{tt}~&=&~\frac{2M \bar{r}}{\bar{r}^2+a^2 \cos^2 \bar{\theta}}
d \bar{t}^2~~, \nonumber \\
\Delta S^2_{t \phi}~&=&~-\frac{4M \bar{r}a \sin^2 \bar{\theta}}
          {\bar{r}^2+a^2 \cos^2 \bar{\theta}} d \bar{t}~d \bar{\phi}~~,
\nonumber \\
\Delta S^2_{rr}~&=&~\frac{2M \bar{r}-a^2 \sin^2 \bar{\theta}}
           {\bar{r}^2-2M \bar{r}+a^2} d \bar{r}^2~~, \nonumber \\
\Delta S^2_{\theta \theta}~&=&~a^2 \cos^2 \bar{\theta}~d \bar{\theta}^2~~,
\nonumber \\
\Delta S^2_{\phi \phi}~&=&~
  \left[ a^2+\frac{2M \bar{r} a^2 \sin^2 \bar{\theta}}
  {\bar{r}^2+a^2 \cos^2 \bar{\theta}} \right] \sin^2 \bar{\theta}
  d \bar{\phi}^2~~. \nonumber
\end{eqnarray}
Here, we discuss two cases: First, for the motion  parallel to the rotational
axis
of the Kerr
black hole, and second, perpendicular. The rotational axis is
taken in the z--direction.

\subsection{Motion Parallel to the Rotational Axis}

First, we investigate a Kerr black hole which moves parallel to the rotational
axis.
As in Sec.II, Lorentz boost of (2.2) is performed for the metric (3.2).
In leading orders of $\gamma$,
the old variables are transformed to
\begin{eqnarray}
d\bar{\theta}~&\longrightarrow&~\frac{-\bar{\rho} du}{u^2+\tilde{\rho}^2}
{}~, \nonumber \\
d\bar{\phi}~&\longrightarrow&~\frac{x~dy-y~dx}{\rho^2}~~, \nonumber
\end{eqnarray}
and
\begin{eqnarray}
\sin \bar{\theta}~&=&~\frac{\tilde{\rho}}{\sqrt{u^2+\tilde{\rho}^2}}~~,
\nonumber \\
\cos \bar{\theta}~&=&~\frac{-u}{\sqrt{u^2+\tilde{\rho}^2}}~~,
\end{eqnarray}
Below, we retain only the terms of $g_{uu} \sim O(\gamma)$.
Because $\lim \Delta S^2_{t \phi} \rightarrow~O(1)$
and $\lim \Delta S^2_{\phi \phi} \rightarrow O(1/\gamma)$,
these terms can be neglected;

$\Delta S^2_{tt}$ of (3.2) is transformed as follows:
\begin{eqnarray}
\Delta S^2_{tt}~\longrightarrow~\frac{ 2p \sqrt{\left(u^2+\tilde{\rho}^2
\right)^3} }
    {u^4+ \left( 2 \tilde{\rho}^2+\tilde{a}^2 \right)
u^2+\tilde{\rho}^4}~du^2~.
\end{eqnarray}
where we put $\tilde{a}=a/\gamma$.
This is integrated using (B1) of Appendix B, then the
limit $\gamma \longrightarrow \infty$ is taken, and finally, it is
differentiated.
The result is
\begin{eqnarray}
\lefteqn{\lim_{\gamma \rightarrow \infty} \Delta S_{tt}^2~\longrightarrow}
&& \nonumber\\
&&~~2p \left[ \frac{1}{|u|}- \left[ \log \rho^2+ \frac{1}{\sqrt{2} \alpha}
  \{\sqrt{\alpha +1} \log \frac{\sqrt{\alpha +1}+\sqrt{2}}{\sqrt{\alpha +1}
  -\sqrt{2}}
    \right. \right. \nonumber\\
&& ~~~~~~~~~~~ \left. \left. -2 \sqrt{\alpha -1} \tan^{-1}
\sqrt{\frac{2}{\alpha-1}} \} \right] \right]
  ~\delta \left(u \right) du^2~~.
\end{eqnarray}
where
\begin{eqnarray}
\alpha = \frac{\sqrt{4 \rho^2 +a^2}}{a}~~.\nonumber
\end{eqnarray}

On the other hand, $\Delta S_{rr}^2$ is converted into the following two
divided terms by the Lorentz transformation:
\begin{eqnarray}
\Delta S_{rr}^2~&\longrightarrow&~\frac{2p~u^2~~du^2}
  {\left[ u^2+\tilde{\rho}^2+\tilde{a}^2-2 \tilde{M} \sqrt{u^2+\tilde{\rho}^2}
   \right] \sqrt{u^2+\tilde{\rho}^2} } \nonumber \\
 &-& \frac{a^2 \tilde{\rho}^2 u^2~~du^2}
  {\left[ u^2+\tilde{\rho}^2+\tilde{a}^2-2\tilde{M} \sqrt{u^2+\tilde{\rho}^2}
   \right] \left( u^2+\tilde{\rho}^2 \right)^2 }~~, \nonumber \\
 &=&~\Delta S_{rr1}^2 +\Delta S_{rr2}^2~~,
\end{eqnarray}
where we put $\tilde{M}=M/\gamma$. Applying (B2), we find the final form of
 $\Delta S_{rr1}^2$ by the same procedures:
\begin{eqnarray}
\lim_{\gamma \longrightarrow \infty} \Delta S_{rr1}^2~\longrightarrow~
2p \left[ \frac{1}{|u|}- \left[ \log \rho^2 -\frac{4}{\beta -\zeta}
 \left[ \sqrt{\beta} \left(1+\zeta \right) \tan^{-1}
\left( 1/\sqrt{\beta} \right) \right. \right. \right.\nonumber\\
-\left. \left. \left. \sqrt{\zeta }\left(1+\beta \right)
\tan^{-1} \left( 1/\sqrt{\zeta} \right) \right]
 \right] \delta \left( u \right) \right] du^2,\nonumber\\
\end{eqnarray}
where
\begin{eqnarray}
\beta =\frac{\rho^2-a^2+2 \rho \sqrt{M^2-a^2}}{\rho^2+a^2+2M \rho}~~,
\nonumber\\
\zeta =\frac{\rho^2-a^2-2 \rho \sqrt{M^2-a^2}}{\rho^2+a^2+2M \rho}~~.
\nonumber
\end{eqnarray}
Similarly, $\Delta S_{rr2}^2$ becomes:
\begin{eqnarray}
\lim_{\gamma \longrightarrow \infty} \Delta S_{rr2}^2\longrightarrow~
-&&\frac{4p \rho}{M a^4} \left[ \left( a^4+2a^2 \rho^2-8\rho^2 M^2 \right)
  -a^2 M\rho \right. \nonumber\\
&&+\left. \frac{1}{4 \sqrt{M^2-a^2}}
\left( \left[\eta \sqrt{\beta}-\theta / \sqrt{\beta} \right]
\tan^{-1} \left( 1/\sqrt{\beta} \right) \right. \right. \nonumber\\
 &&~\left. \left. -\left[ \eta \sqrt{\zeta}-\theta /\sqrt{\zeta} \right]
\tan^{-1}
\left(1/\sqrt{\zeta} \right) \right) \right]~\delta \left( u \right)~du^2~~,
\nonumber\\
\end{eqnarray}
where
\begin{eqnarray}
\eta~=~\left( 4\rho M^2-a^2 \rho-2 a^2 M \right)
\left( \rho^2+a^2+2 \rho M \right)~~, \nonumber\\
\theta~=~\left( 4\rho M^2-a^2 \rho+2 a^2 M \right)
\left( \rho^2+a^2-2 \rho M \right)~~, \nonumber
\end{eqnarray}
By the Lorentz transformation, the $\Delta S_{\theta \theta}^2$ becomes
\begin{eqnarray}
\Delta S_{\theta \theta}^2~\longrightarrow~\frac{a^2 \tilde{\rho}^2 u^2~~du^2}
  {\left( u^2+\tilde{\rho}^2 \right)^3}~~,
\end{eqnarray}
and using (B4), after taking the limit $\gamma \rightarrow \infty$ is:
\begin{eqnarray}
\lim_{\gamma \longrightarrow \infty} \Delta S_{\theta \theta}^2
4p \left( \frac{ \pi a^2}{32 \rho M} \right) \delta \left( u \right)~du^2~~.
\end{eqnarray}
Putting (3.4) $\sim$ (3.10) altogether, and carrying out the coordinate
transformation, the exact gravitational shock wave solution for a Kerr black
hole
moving parallel to the rotational axis is given by
\begin{eqnarray}
\lim_{\gamma \to \infty} ds_K^2~\longrightarrow~&&-du~dv+dx^2+dy^2
\nonumber\\
&&-4p \left[ \Delta g_1+\Delta g_2+\Delta g_3+\Delta g_4 \right]~
\delta(u)~du^2~~,
\end{eqnarray}
where
\begin{eqnarray}
\Delta g_1& = &\log \rho^2+\frac{1}{2\sqrt{2} \alpha}
  \left[ \sqrt{\alpha +1} \log \frac{\sqrt{\alpha +1} +\sqrt{2}}
       {\sqrt{\alpha +1} -\sqrt{2}} \right. \nonumber\\
  &&~~~\left. -2 \sqrt{\alpha -1} \tan^{-1} \sqrt{\frac{2}{\alpha-1}} \right]
  ~~, \nonumber\\
\Delta g_2 &= &-\frac{2}{\beta -\zeta}
 \left[ \sqrt{\beta} \left(1+\zeta \right) \tan^{-1}
\left(1/\sqrt{\beta} \right) -\sqrt{\zeta} \left(1+\beta \right)
\tan^{-1} \left(1/\sqrt{\zeta} \right) \right], \nonumber\\
\Delta g_3 &=&
  \frac{\rho}{M a^4} \left[ \left( a^4+2a^2 \rho^2-8\rho^2 M^2 \right)~\pi/8
  -a^2 M\rho \right. \nonumber\\
&&~~~+\left. \frac{1}{4 \sqrt{M^2-a^2}} \left[ \left[\eta \sqrt{\beta}-\theta
/ \sqrt{\beta} \right]
\tan^{-1} \left(1/\sqrt{\beta} \right) \right. \right. \nonumber\\
&&~~~~~\left. \left. -\left[ \eta \sqrt{\zeta}-\theta /\sqrt{\zeta} \right]
\tan^{-1}
\left(1/\sqrt{\zeta} \right) \right] \right]~~, \nonumber\\
\Delta g_4 &=& -\frac{\pi a^2}{32 \rho M}~~. \nonumber
\end{eqnarray}

Assume there are no naked singularities. Then a black
hole with $a=M$ is the maximally rotating black hole.
We now investigate the properties of (3.11) in the limit of
$M \rightarrow a$.
In this limit, (3.11) has the form:
\begin{eqnarray}
\lim_{M \rightarrow a}&& ds^2_K \rightarrow -du~dv+dx^2+dy^2 \nonumber\\
 && -4p \left[ \log \rho^2 +1+\frac{1}{2 \sqrt{2}} \frac{1}{\alpha}
 \left( \sqrt{\alpha +1} \log \frac{\sqrt{\alpha+1}+\sqrt{2}}{\sqrt{\alpha+1}
 -\sqrt{2}}
    \right. \right. \nonumber\\
   &&\left. \left. -2 \sqrt{\alpha-1} \tan^{-1} \sqrt{\frac{2}{\alpha-1}}
\right)
 +\frac{ \left( 3\rho^4 -2a^2 \rho^2-2a^4 \right)}{a^3 \sqrt{\rho^2-a^2}}
   \tan^{-1} \sqrt{\frac{\rho+a}{\rho-a}} \right.\nonumber\\
&&~~ +\left. \left( \frac{\rho}{8a}-\frac{3 \rho^3}{4a^3}-\frac{a}{32 \rho}
\right) \pi
-\frac{3 \rho^2}{2a^2} \right]~\delta \left( u \right) du^2~~.
\end{eqnarray}

The shift $\Delta v$ for $\rho \sim a$ is shown in Fig.2 by the thin solid
line.
Notice that the shift is divergent at $\rho = a$,
the event horizon of a Kerr black hole for $M \rightarrow a$.
On the other hand, when $\rho \gg a$
\begin{eqnarray}
\lim_{M \longrightarrow a}&& ds^2_K {\longrightarrow \atop \rho \gg a}
-du~dv+dx^2+dy^2 \nonumber\\
&&-4p \left[ \log \rho^2+1-\frac{\pi}{2} \frac{a}{\rho}+\left( \frac{4}{3}
-\frac{4}{5} \right)
 \frac{a^2}{\rho^2}+O\left( \frac{a^3}{\rho^3} \right) \right]~
  \delta \left( u \right)~du^2~~. \nonumber\\
\end{eqnarray}
Compared with (2.13) of Schwarzschild type, (3.13) is slightly different in
the order of
$O \left( a^2/\rho^2 \right)$.
Thus, far from the hole, the metric of the Kerr type hole moving in
the z--direction is not distinguished
from that of the Schwarzschild type one.

The refraction angle derived from (3.13) is presented in Fig.3 by
the thin solid line.
As mentioned in Sec.II, the refraction angle is 90${}^{\circ}$ at $\rho = M$,
 the event horizon of the Kerr black hole.

\subsection{The Motion Perpendicular to the Rotational Axis}

In this section we study the metric in a reference system  moving
perpendicular to the rotational axis of a Kerr black hole.
The moving direction is taken as the x--direction, and
the rotational axis is set as the z--direction. We discuss only in the
equatorial
plane
($\theta = \pi/2,~\dot{\theta} =0$).

The boost in the x--direction at the speed $v$ gives:
\begin{eqnarray}
t=\gamma \left( \bar{t}+v \bar{x} \right),~x= \gamma \left(\bar{x}
+v \bar{t} \right),~
y=\bar{y},
\end{eqnarray}
where
\begin{eqnarray}
\gamma~=~(1-v^2)^{-1/2} ~~,\nonumber
\end{eqnarray}
and z is always set to zero.
We put $u=t-x$ and $v=t+x$ as the null coordinates.
Then,  old coordinates are transformed as follows:
\begin{eqnarray}
&&\bar{t} \rightarrow \gamma u, \nonumber\\
&&\bar{x}~\rightarrow~-\gamma u, \nonumber\\
&&\bar{r}^2~\longrightarrow~\gamma^2 u^2+y^2,
\end{eqnarray}
and the infinitesimal lengths, $d\bar{r}$ and $d\bar{\phi}$, are transformed as
( in
the leading order of $\gamma$) :
\begin{eqnarray}
d \bar{r}~&\longrightarrow &~\frac{\gamma u~du}{\sqrt{u^2+\tilde{y}^2}} ,
 \nonumber\\
d \bar{\phi}~&\longrightarrow &~\frac{1}{\gamma} \frac{y~du}{u^2
+\tilde{y}^2}~~.
\end{eqnarray}

Making use of (3.15) and (3.16), (3.2) is changed under the Lorentz
transformation as:
\begin{eqnarray}
ds^2_K \rightarrow -du~dv &+& dy^2 \nonumber\\
&+&\left( \frac{2p}{\sqrt{u^2+\tilde{y}^2}} -\frac{1}{\gamma}
\frac{4May}{\sqrt{\left(u^2+\tilde{y}^2 \right)^3}}
\right.\nonumber\\
&+&\left. \frac{2pu^2}{\left[ u^2+\tilde{y}^2+\tilde{a}^2-2\tilde{M}
\sqrt{u^2+\tilde{y}^2} \right]
\sqrt{u^2+\tilde{y}^2}} \right. \nonumber\\
&-& \left. \frac{a^2~u^2}{\left[ u^2+\tilde{y}^2+\tilde{a}^2-2\tilde{M}
\sqrt{u^2+\tilde{y}^2} \right]
\left( u^2+\tilde{y}^2 \right) } \right.\nonumber\\
&+& \left. \frac{2pa^2 y^2}{\gamma^4 \left( u^2+\tilde{y}^2+\tilde{a}^2
\right)
  \sqrt{ \left( u^2+\tilde{y} \right)^3 } } \right)~du^2~~. \nonumber\\
\end{eqnarray}
Now we take the limit $\gamma \rightarrow \infty$. Using the integration
formulae of
(3.2), (B2), (B5), (B6) and (B7),
the metric calculated by the same method of Sec.II and Sec.IIIa becomes:
\begin{eqnarray}
\lim_{\gamma \to \infty} ds^2_K~\longrightarrow~&&-du~dv+dy^2
\nonumber\\
&&-4p \left[ \delta g_1+\delta g_2+\delta g_3+\delta g_4 \right]~
\delta(u)~du^2~~,
\end{eqnarray}
where
\begin{eqnarray}
\delta g_1~&&=~\log |\tilde{y}|^2-\frac{2}{\beta^{\prime}-\zeta^{\prime}}
\left( \sqrt{\beta^{\prime}}
  \left( 1+\zeta^{\prime} \right) \tan^{-1} \left( 1/\sqrt{\beta^{\prime}}
  \right)\right. \nonumber\\
 &&\left. -\sqrt{\zeta^{\prime}} \left( 1+\beta^{\prime} \right)
	 \tan^{-1} \left( 1/\sqrt{\zeta^{\prime}} \right) \right)~~, \nonumber\\
&&~~~~\beta^{\prime}~=~\frac{y^2-a^2+2 |y| \sqrt{M^2-a^2}}{y^2+a^2
+2M |y|}~~, \nonumber\\
&&~~~~\zeta^{\prime}~=~\frac{y^2-a^2-2 |y| \sqrt{M^2-a^2}}
{y^2+a^2+2M |y|}~~,\nonumber\\
\delta g_2~&&=~\frac{2 |y| a^2}{M \left\{|y| \left(\sqrt{M^2-a^2}+M \right)
+a^2 \right\} }~~, \nonumber\\
 &&~~~~\times~\left\{ \frac{1}{\beta^{\prime}-1}
  \left[ \frac{\pi}{4}-\frac{1}{\sqrt{\beta^{\prime}}} \tan^{-1}
  \left( 1/\sqrt{\beta^{\prime}} \right) \right] \right. \nonumber\\
 &&~~~~ \left. +\frac{\zeta^{\prime}}{\zeta^{\prime}-\beta^{\prime}}
  \left[ \frac{1}{\sqrt{\zeta^{\prime}}} \tan^{-1} \left(
1/\sqrt{\zeta^{\prime}}
  \right)
     -\frac{1}{\sqrt{\beta^{\prime}}} \tan^{-1} \left( 1/\sqrt{\beta^{\prime}}
	 \right) \right]
  \right\} \nonumber\\
\delta g_3~&&=~\frac{2a}{y} \nonumber\\
\delta g_4~&&=~-1-\frac{y^2}{2a \sqrt{y^2+a^2}} \log
\frac{\sqrt{y^2+a^2}-a}{\sqrt{y^2+a^2}+a}~~. \nonumber
\end{eqnarray}
Here, the important point to be noted is that $\delta g_3$ depends on the sign
of $y$ explicitly.
In other words, when the magnitudes of $y$'s are equal with opposite signs,
the shift $\Delta v$'s and the refraction angles are different.
As the term is derived from $g_{t \phi} dt d\phi$ of the Kerr metric,
the effect of the space--time dragging would be the cause.

(3.18) is the general result. Now we discuss the properties of it for special
cases.
Take the limit $M \rightarrow a$;  i.e. the fast rotating black hole:
\begin{eqnarray}
\lim_{M \longrightarrow a} &&\left[ \lim_{\gamma \longrightarrow \infty}
ds^2_K \right]
 ~\longrightarrow ~-du~dv+dy^2 \nonumber\\
&&-4p \left[ \log |y|^2 -\frac{1}{2}+\frac{y^2-2a^2}{a \sqrt{y^2-a^2}}
   \tan^{-1} \sqrt{\frac{|y|+a}{|y|-a}} -\frac{|y|}{4a} \pi \right. \nonumber\\
 && \left. +\frac{2a}{y}-\frac{y^2}{2a \sqrt{y^2+a^2}} \log
 \frac{\sqrt{y^2+a^2}-a}{\sqrt{y^2+a^2}+a}
 \right] \delta \left( u \right)~ du^2~~.
\end{eqnarray}

Consider the case of $|y|>>a$, far from the hole:
\begin{eqnarray}
&&ds^2_K {\longrightarrow \atop |y| \gg a} -du~dv+dy^2 \nonumber\\
   &&~-4p \left[ \log |y|^2+1-\frac{3 \pi}{8} \left( \frac{a}{|y|} \right)
   +2\left( \frac{a}{y} \right)
  -\frac{4}{3} \left( \frac{a^2}{y^2} \right) +O\left( \frac{a^3}{y^3} \right)
  \right]
     \delta \left( u \right)~du^2. \nonumber\\
\end{eqnarray}
On the other hand, in the case of $|y| \sim a$, the shift $\Delta v$ is shown
in
Fig.2.

The refraction angle is depicted in Fig.3.
In Fig.4, the ratio of the refraction angle of the negative sign of $y$ to that
of
the positive sign
is represented.
This figure indicates that the refraction for $y<0$ is larger than
that for $y>0$.

\setcounter{equation}{0}
\section{Astrophysical Applications}

In this section, we discuss shortly possible applications of our results to
some
astrophysical
phenomena.
The metrics obtained above are the first--order approximate equations of
Lorentz factor
$\gamma$.
Therefore, the metrics are exact in the limit $\gamma \rightarrow \infty$.
If the $\gamma$ factor of an object is $\sim$ 10, the metric of the
gravitational
shock wave is
applied to the object with sufficient reliability.
Quasars which are observed in recent years have, however, $\sim$ 2--3,
 so that discussions for quasars below would be qualitative in nature.

The metric can be applied to the problem of gravitational waves
which are radiated by the collision of two black holes
and the scattering problem of particle physics
at energies of Planck scale. These problems have also been calculated by using
the A--S
metric \cite{D'Eath92,'t Hooft87}. It will be very interesting when the result
calculated
by our metric is
compared with that by the A--S metric, because our case can be traced even
for the region of
$\rho \sim M$. This is now under investigations, and will be published in
the next paper.

Let us return to our main subject. The main point of us compared with others is
that black holes are moving.

Fig.5 shows the effect of the gravitational lens.
The middle object represents a supermassive black hole, and the left side
object is
the light source (i.e. a quasar for example). The observer at the right will
see
 two or many images, because the light rays are refracted near the black hole.
The light source is, in fact, receding with a Lorentz factor $\gamma$
from us caused by the expansions of the universe.

Let us discuss in the rest frame of the source.
In this frame, the black hole moves in the right direction with
the Lorentz factor $\gamma$.
For a Schwarzschild black hole, the light radiated from the source is refracted
by the angle $\Delta\hat{ \phi}_S$ (2.13):
\begin{eqnarray}
\Delta\hat{ \phi}_S~\sim~\frac{4p}{\rho_0}
  \left[ 1+\frac{\pi}{2} \frac{M}{\rho_0}-\frac{8}{3} \frac{M^2}{\rho^2_0}
     +O \left( \frac{M^3}{\rho^3_0} \right) \right]~~,
\end{eqnarray}
when $\rho_0 >> M$. In the observer's frame, the refraction angle
$\Delta \phi_S$ is given from
$\Delta\hat{ \phi}_S$ by the Lorentz transformation as
 $\Delta \phi_S \sim \Delta\hat{ \phi}_S/\gamma$.
Then, $\Delta \phi_S$ is given by:
\begin{eqnarray}
\Delta \phi_S~\sim~\frac{4M}{\rho_0}
  \left[ 1+\frac{\pi}{2} \frac{M}{\rho_0}-\frac{8}{3} \frac{M^2}{\rho^2_0}
     +O \left( \frac{M^3}{\rho^3_0} \right) \right]~~,
\end{eqnarray}
We would like to emphasize that $p$ of (2.13) is replaced by $M$.
The naive value of  $\Delta \phi_S$ is $\sim 4M/\rho_0$, so that the effect of
$\gamma$ appears in the order of $M^2/\rho^2_0$.

Similarly, when the black hole is of the Kerr type, we can calculate
the refraction angle,
$\Delta \phi_{Kz}$, and $\Delta \phi_{Kx}$, of the parallel motion to
the rotational axis from (3-14),
and of the perpendicular motion from (3.21), respectively:
\begin{eqnarray}
\Delta \phi_{Kz}~\sim~\frac{4M}{\rho_0}
  \left[ 1+\frac{\pi}{2} \frac{M}{\rho_0}-\left(\frac{8}{3} -\frac{8}{5}
  \right)
      \frac{M^2}{\rho^2_0}
     +O \left( \frac{M^3}{\rho^3_0} \right) \right]~~,\\
\Delta \phi_{Kx}~\sim~\frac{4M}{y_0}
  \left[ 1+\left( \frac{3 \pi}{8}-2 \right) \frac{a}{y_0}
    +\frac{8}{3} \frac{M^2}{y^2_0}
	+O \left( \frac{M^3}{y^3_0} \right) \right]~~.
\end{eqnarray}

On the other hand, in the near region  $\rho_0 \sim a$ or $y_0 \sim a$,
$p$ of the vertical axis in Fig.3  is to be replaced by $M$ as discussed above.

\section{Conclusion and Discussion}

We have derived the metrics of a moving Schwarzschild black hole and
a moving Kerr black hole by
Lorentz transformations. The Schwarzschild metric is calculated in
the ordinary coordinates,
not in the isotropic coordinates, and the Kerr metric in the Boyer--Lindquist
coordinates.
For the Kerr black hole, we have calculated the metrics for two cases:
the parallel and the
perpendicular cases.
For both cases, obtained metrics are plane--fronted shock waves and
the geometries of them
have structures of two portions
of the Minkowski types for $u<0$ and $u>0$, with null shock surfaces at
$u=0$ with warps. The metrics can be investigated in the region near
the black hole
at $u=0$, because we do not put $M \rightarrow 0$ which was required for
the A--S metric.
Next, we have discussed the shift, $\Delta v$, of a null coordinate $v$
crossing
$u=0$, and
have obtained the refraction angle from our metric.

Our main results are:
\newcounter{mycount}
\begin{list}{}{\setlength{\leftmargin}{3em}
  \setlength{\labelsep}{0em}
  \setlength{\labelwidth}{0pt}
  \usecounter{mycount}
  \renewcommand{\makelabel}{(\roman{mycount})}
}
\item The exact formulae for the metrics of Schwarzschild and
Kerr black holes moving
with the Lorentz factors $\gamma \rightarrow \infty$ are obtained.
\item The shift $\Delta v$'s of the null coordinates crossing
the gravitational shock at $u=0$
for a Kerr black hole is derived, both for parallel and perpendicular motions.
\item The refraction angle of null geodesics passing by
a Kerr black hole is obtained and
it is found that it depends on the sign of $y$.
\item The gravitational lens effect by a moving black holes manifests in
the second order
of $M/\rho$, compared with static cases.
\end{list}

This is the first step to the physics of moving black holes.
Nature waits for our endeavors to look into her mysterious secretes !

\clearpage

\renewcommand{\theequation}{\Alph{section}\arabic{equation}}
\appendix
\renewcommand{\thesection}{Appendix \Alph{section}: }

\section{Geodesics of the metric  (1.1)}
\setcounter{equation}{0}

For the metric (1.1), we put $x=\rho \cos \phi$ and $y=\rho \sin \phi$.
Then, it has the following form:
\begin{eqnarray}
\lim_{\gamma \to \infty} ds^2~\longrightarrow~-du~dv+d\rho^2+\rho^2~
d\phi^2-A(\rho)~\delta(u)~du^2~~.
\end{eqnarray}
The null geodesics derived from (A1) satisfy:
\begin{eqnarray}
&&\ddot{u}~=~0~~, \\
&&\frac{d}{d\lambda} \left( \rho^2 \dot{\phi} \right)~=~0~~, \\
&&\frac{d}{d\lambda} \left( \dot{v}+2 A \left( \rho \right) \dot{u} \delta
   \left( u \right) \right)~=~0~~, \\
&&\ddot{\rho}-\rho \dot{\phi}^2+\frac{1}{2} A' \left( \rho \right) \dot{u}^2
\delta \left( u \right)~=~0~~,
\end{eqnarray}
where dash and dot denote the derivatives with respect to $\rho$ and the affine
parameter $\lambda$, respectively.

{}From (A2) we can put
\begin{eqnarray}
u~=~\lambda~~,
\end{eqnarray}
without loss of generality. Then, (A4) means
\begin{eqnarray}
\left. v~=~-2 A \left( \rho \right) \right|_{\rho=\rho_0} \theta
\left( u \right)~~,
\end{eqnarray}
where inessential constants have been neglected. The shift $\Delta v$ of
the geodesic
$v$ crossing $u=0$ is now given by
\begin{eqnarray}
\left. \Delta v~=~-2 A \left( \rho \right) \right|_{\rho=\rho_0} ~~,
\end{eqnarray}
(A3) represents the conservation of the angular momentum:
\begin{eqnarray}
\rho^2 \dot{\phi}~=~{\rm{constant}}~\equiv L~~.
\end{eqnarray}
(A5) is written by now as:
\begin{eqnarray}
\ddot{\rho}-\frac{L^2}{\rho^3}+\frac{1}{2} A' \left( \rho \right)
\delta \left( u \right)~=~0~~.
\end{eqnarray}
In (A10), if the last term were absent, it would give the straight geodesic,
$\rho \cos \phi=\rho_{0}$.
Since we are interested in the behavior of the geodesic crossing the shock
$u=0$, we neglect the second term of (A10). The solution is now given
by:
\begin{eqnarray}
\left. \dot{\rho}~=~- \frac{1}{2} A' \left( \rho \right) \right|_
{\rho =\rho_{0}}
\theta \left( u \right)~~.
\end{eqnarray}
{}From (A11), the refraction angle $\Delta \phi$ is given:
\begin{eqnarray}
\left. \tan \left( \Delta \phi \right)~=~\frac{1}{2} \frac{d A
\left( \rho \right)}{d \rho}
 \right|_{\rho=\rho_0}~~.
\end{eqnarray}

\section{Integration formulae}

\setcounter{equation}{0}
\begin{eqnarray}
\int && \frac{\sqrt{\left( u^2+\tilde{\rho}^2 \right)^3}~du}
   {u^4+\left( 2 \tilde{\rho}^2+\tilde{a}^2 \right) u^2+\tilde{\rho}^4}
   \nonumber\\
&&=~\log \frac{\sqrt{u^2+\tilde{\rho}^2}+u}{\tilde{\rho}}
 -\frac{1}{2 {\cal{A}} } \left\{ \sqrt{ \frac{ {\cal{A}}+1}{2}}
 \log \frac{\sqrt{ \left( {\cal{A}}+1 \right) \left(u^2+\tilde{\rho}^2 \right)
}
 +\sqrt{2}u }
         {\sqrt{ \left( {\cal{A}}+1 \right) \left(u^2+\tilde{\rho}^2 \right) }-
		 \sqrt{2}u } \right. \nonumber\\
&&\left. -2 \sqrt{ \frac{ {\cal{A}}-1}{2}} \tan^{-1}
      \frac{ \sqrt{2} u}{\sqrt{ \left( {\cal{A}}-1 \right) \left( u^2+
	  \tilde{\rho}^2 \right) }} \right\}~~,
\end{eqnarray}
where tilde denotes the division by $\gamma$, and
\begin{eqnarray}
{\cal{A}}=\frac{ \sqrt{4 \rho^2+a^2}}{a}~~. \nonumber
\end{eqnarray}
\begin{eqnarray}
 \int && \frac{u^2~du}
       {\left[ u^2+\tilde{\rho}^2+\tilde{a}^2-2\tilde{M} \sqrt{u^2+
	   \tilde{\rho}^2} \right]
          \sqrt{u^2+\tilde{\rho}^2} } \nonumber\\
&&=~\log \frac{1+q}{1-q} \nonumber\\
&&+\frac{2}{ {\cal{B}}-{\cal{C}}}
   \left[ \sqrt{{\cal{B}} }\left( 1+{\cal{C}} \right) \tan^{-1}
   \left( q/\sqrt{ {\cal{B}}} \right)
        - \sqrt{{\cal{C}}} \left( 1+{\cal{B}} \right) \tan^{-1}
		\left( q/\sqrt{ {\cal{C}}} \right) \right]~~,
              \nonumber\\
\end{eqnarray}
where
\begin{eqnarray}
&&q~=~\frac{\sqrt{u^2+\tilde{\rho}^2}-\tilde{\rho}}{u}~~, \nonumber\\
&&{\cal{B}}~=~\frac{\rho^2-a^2+2 \rho \sqrt{M^2-a^2}}
{\rho^2+a^2+2 M\rho}~~,\nonumber\\
&&{\cal{C}}~=~\frac{\rho^2-a^2-2 \rho \sqrt{M^2-a^2}}
{\rho^2+a^2+2 M\rho}~~.\nonumber
\end{eqnarray}
\begin{eqnarray}
\int &&  \frac{u^2~~du}
{ \left[ u^2+\tilde{\rho}^2+\tilde{a}^2-2 \tilde{M}
   \sqrt{u^2+\tilde{\rho}^2} \right] \left( u^2+\tilde{\rho}^2 \right)^2 }
   \nonumber\\
 &&=~\frac{4 \gamma^3}{\rho \left( \rho^2+a^2+2M \rho \right) }
\left[ {\cal{D}} \tan^{-1}q+{\cal{E}} \frac{q}{1+q^2} \right. \nonumber\\
       && \left. +{\cal{F}} \frac{q}{\left( 1+q^2 \right)^2}
	 +~\frac{1}{{\cal{B}} -{\cal{C}}}
		 \left\{ \left[ {\cal{G}} \sqrt{{\cal{B}}}-{\cal{H}}/\sqrt{{\cal{B}}} \right]
       \tan^{-1} \left( q/\sqrt{{\cal{B}}} \right) \right. \right. \nonumber\\
		  && \left. \left. - \left[ {\cal{G}} \sqrt{{\cal{C}}}-{\cal{H}}/
		  \sqrt{{\cal{C}}} \right]
       \tan^{-1} \left( q/\sqrt{{\cal{C}}} \right) \right\} \right],
\nonumber\\
\end{eqnarray}
where $q$, ${\cal{B}}$, and ${\cal{C}}$ are same as (B2),
\begin{eqnarray}
{\cal{D}}~&=&~\left( a^4+2 a^2 \rho^2-8\rho^2 M^2 \right)
              \left( \rho^2+a^2+2 \rho M  \right) /4 a^6~~,\nonumber\\
{\cal{E}}~&=&~\left( a^2-4 M \rho \right) \left( \rho^2+a^2+
2 \rho M  \right) /4 a^4~~,\nonumber\\
{\cal{F}}~&=&~-\left( \rho^2+a^2+2 \rho M  \right) /2 a^2~~, \nonumber\\
{\cal{G}}~&=&~\rho \left( 4 \rho M^2-a^2 \rho -2 a^2 \rho^2 \right)
         \left( \rho^2+a^2+2 \rho M  \right)/2a^6~~, \nonumber\\
{\cal{H}}~&=&~\rho \left( 4 \rho M^2-a^2 \rho +2 a^2 \rho^2 \right)
         \left( \rho^2+a^2-2 \rho M  \right)/2a^6~~. \nonumber
\end{eqnarray}
\begin{eqnarray}
\int && \frac{u^2~~du}{ \left( u^2+\tilde{\rho}^2 \right)^3 } \nonumber\\
&&=~\frac{1}{8}
  \left\{ \frac{1}{\tilde{\rho}^3} \tan^{-1} \left( u/\tilde{\rho} \right)
   +\frac{1}{\tilde{\rho}^2} \frac{u}{u^2+\tilde{\rho}^2}
      -\frac{2u}{\left( u^2+\tilde{\rho}^2 \right)^2} \right\} ~~.~~~~~~~~~
\end{eqnarray}
\begin{eqnarray}
\int \frac{du}{\sqrt{\left( u^2+\tilde{y}^2 \right)^3 }}
  ~=~\frac{u}{\tilde{y}^2 \sqrt{u^2+\tilde{y}^2}}~~.
  ~~~~~~~~~~~~~~~~~~~~~~~~~~~~~~~~
\end{eqnarray}
\begin{eqnarray}
\int &&\frac{u^2~~du}
  {\left[ u^2+\tilde{y}^2+\tilde{a}^2-2 \tilde{M} \sqrt{u^2+\tilde{y}^2}
\right]
     \left( u^2+\tilde{y}^2 \right) } \nonumber\\
&&=~\frac{4 \gamma |y|}{a^2+|y| \left( M+\sqrt{M^2-a^2} \right) }
\nonumber\\
   &&\left\{ \frac{1}{{\cal{J}} -1}
        \left[ \tan^{-1} w-\frac{1}{\sqrt{{\cal{J}}}}
           \tan^{-1} \left( w/ \sqrt{{\cal{J}}} \right) \right] \right.
 \nonumber\\
&&\left. -\frac{{\cal{K}}}{{\cal{J}}-{\cal{K}}}
  \left[ \frac{1}{\sqrt{ {\cal{K}}}} \tan^{-1} \left( w/ \sqrt{{\cal{K}}}
\right)
    -\frac{1}{\sqrt{{\cal{J}}}} \tan^{-1} \left( w/ \sqrt{{\cal{J}}} \right)
	\right] \right\}~~,
\end{eqnarray}
where
\begin{eqnarray}
w~&=&~\frac{\sqrt{u^2+\tilde{y}^2}-|\tilde{y}|}{u}~~, \nonumber\\
{\cal{J}}~&=&~\frac{y^2-a^2+2 |y| \sqrt{M^2-a^2}}{y^2+a^2+2 M |y|}~~,
\nonumber\\
{\cal{K}}~&=&~\frac{y^2-a^2-2 |y| \sqrt{M^2-a^2}}{y^2+a^2+2 M |y|}~~.
\end{eqnarray}
\begin{eqnarray}
\int && \frac{du}{ \left( u^2+\tilde{y}^2+\tilde{a}^2 \right)
    \sqrt{ \left( u^2+\tilde{y}^2 \right)^3}} \nonumber\\
&&=~\frac{1}{\tilde{y}^2 \tilde{a}^2}
  \left\{ \frac{u}{ \sqrt{u^2+\tilde{y}^2}}+\frac{y^2}{2 a \sqrt{y^2+a^2}}
     \log \frac{ \sqrt{y^2+a^2} \sqrt{u^2+\tilde{y}^2}-au}
           { \sqrt{y^2+a^2} \sqrt{u^2+\tilde{y}^2}+au} \right\}~~. \nonumber\\
\end{eqnarray}

\clearpage

{\large\bf{Figure Captions}}
\vspace{1em}

\newcounter{mycount1}
\begin{list}{}{\setlength{\leftmargin}{1em}
  \setlength{\labelsep}{0em}
  \setlength{\labelwidth}{0pt}
  \usecounter{mycount1}
  \renewcommand{\makelabel}{Fig.\arabic{mycount1} :}
}
\item The refraction angle, $\Delta \phi$, of a null geodesic when the light is
injected perpendicular to the shock wave.
(a) the Schwarzschild case and the Kerr case moving parallel to the rotational
axis.
The black hole is moving in the $z$--direction.
(b) the Kerr case moving perpendicular to the axis. The black hole is moving in
the $x$--direction.
An arrow of the circle near the center represents the rotational direction of
the Kerr black hole.
\item The shift $\Delta v$ crossing $u=0$. We put $M=1$.
The bold line indicates the Schwarzschild case, and the thin solid line
indicates
the Kerr case moving parallel to the
rotational axis, where the limit $M \rightarrow a$ is taken. The dashed line
and
the dotted line represent the Kerr case moving perpendicular
to the axis, where the limit $M \rightarrow a$ is taken. The dashed line
indicates
$y >0$,
and the dotted line indicates $y < 0$.
\item The refraction angle, $\Delta \phi$, crossing $u = 0$ where the light ray
is injected perpendicular
to the shock wave.
Notations are the same as in Fig.2.
\item The ratio of the refraction angle of the negative sign of $y$ to that of
the positive sign.
\item The effect of the gravitational lens.
\end{list}

\end{document}